\documentclass[draft,published,notoc]{JHEP} 

\usepackage{epsfig}

\newcommand\fverb{\setbox\pippobox=\hbox\bgroup\verb}
\newcommand\fverbdo{\egroup\medskip\noindent%
			\fbox{\unhbox\pippobox}\ }
\newcommand\fverbit{\egroup\item[\fbox{\unhbox\pippobox}]}
\newbox\pippobox

\title{Enhanced Structure Function Of The Nucleon From Deep Inelastic
Electron-Proton Scattering}

\author{ N. M. Hassan, W. R. El-Harby, R.W.El-Moualed And  M.T. Hussein \\ 
 Physics Department, Faculty of Science, Cairo University, Cairo, Egypt 12613\\ 
        E-mail: \email{hussein1@frcu.eun.eg}}
\received{\today} 		

\preprint{\hep-ph{0001315}}      

\abstract{The proton structure function is re-deduced from the data of deep inelastic
electron-proton scattering after enhanced correction that is made due to the
multiple scattering effect. The Glauber approach is used to account for the
multiple scattering of electron with the quark constituents of the proton.
The results are compared with the prediction of the parton model.
Appreciable deviation is observed only at low values of the Bjorken
scaling, which correspond to the extremely deep inelastic scattering.}


\begin{document} 

\maketitle 

\section{Introduction}

A dynamical understanding of quark substructure had its origin from
experiments on deep inelastic lepton-nucleon scattering [1]. These showed
that the complicated production of many hadrons in such a collision could be
simply interpreted as quasi-elastic scattering of the lepton by point like
particles or quarks. For example the linear relation between the total
inelastic cross section $\sigma _{in}^{tot}(e-p)$and the center of mass
energy $E$ may be explained as simple elastic scattering with point particle
which depends on the coupling constant and the phase space factors. It is
well known that at high momentum transfer, the elastic form factor is very
small and the inelastic scattering of the incident electron is much more
probable than the elastic scattering. In this case there must be an extra
variable $x$ (the Bjorken scaling variable) which relates the momentum
transfer $q$ to the energy transfer $\nu $ by the relation $x=\frac{q^{2}}{%
2M\nu }\ ,0<x<1$ . The case of $x=1$ corresponds to the elastic scattering,
and it tends to zero as the reaction goes toward the deep inelastic. The
differential cross section [2] is then written as,

\begin{equation}
\frac{d^{2}\sigma }{dq^{2}d\nu }=\frac{4\pi \alpha ^{2}}{q^{4}}\frac{%
E^{\prime }}{E}[F_{2}(q^{2},\nu )\cos ^{2}\frac{\theta }{2}+\frac{2\nu }{M}%
F_{1}(q^{2},\nu )\sin ^{2}\frac{\theta }{2}]  \nonumber
\end{equation}

Where the factor represents the square of the scattering amplitude of the
electron by a point charged particle (the parton or quark). $F_{1}$ and $%
F_{2}$ are the hadron structure functions used to correct for the composite
structure of the target,$E$and $E^{^{\prime }}$ being the incident and
scattered electron energies. The factor $cos^{2}(\frac{q}{2})$stands for the
square of the rotational matrix element for the electric non-spin flip part
of the scattering of the electron by a charged point particle. The factor $%
sin^{2}(\frac{q}{2})$ is the corresponding factor for the spin-flip magnetic
interaction between the magnetic moments of the electron and target
particle. The problem arises now revolves about the factor $4\pi \alpha
^{2}/q^{4}$ , which assumes a simple scattering of the electron by a Coulomb
field of a charged point particle. And since the nucleon bag contains 3-
valence quarks rather than the sea quark-antiquark pairs, so we expect
multiple collision inside this bag instead of a single one. Also the
presence of more than one quark in the system may make shadowing on the
others during their collision. Of course these will affect the shape of the
differential cross section and hence the structure functions $F_{1}$ and $%
F_{2}$. The present work aims to consider the effect of multiple scattering
by using the Glauber [3] approach and then re-deduce the form of the
structure function and compare the results with the previously published
data at different regions of the Feynman variable $x$.

\section{The Multiple Scattering Approach}

Let us consider a bound system of three point particles (valence quarks)
forming a proton as a spherical bag. The sea quark-antiquark pairs are
considered as electric dipoles in their plasma phase. The problem is treated
in the laboratory system in which its origin coincides with the center of
the spherical bag. Assuming a collision axis in the z- direction, an
incident electron with position vector and impact parameter will scatter
with a momentum transfer . The scattering amplitude is given by [4],

\begin{equation}
f_{ms}(q)=\frac{k}{2\pi i}\int d^{2}\overline{b}\exp (i\stackrel{\_}{q}.%
\overline{b})\left\langle \Psi \left\{ s_{i}\right\} \left| \exp (i\Xi
(b,\{s_{i}\})-1\right| \Psi \{s_{i}\}\right\rangle 
\end{equation}

Where, $k^{\prime }$ is the momentum of the scattered electron. In case of
inelastic scattering, only a fraction x of the momentum of the incident
electron will transfer to the quark system of the proton, leaving a momentum
fraction $(1-x)$ to the scattered electron. is the total phase shift
function due to all the constituent valence quarks and the associated
dipoles of the system having position vectors $\{r_{i}\}$ with projections $%
\{s_{i}\}$ on the impact parameter plane. The suffix $i$ runs over all the
flavors forming the proton bag. is the total wave function of the proton
system. In Quantum Chromodynamics [5,6,7], the force between quarks is
mediated by the exchange of massless spin 1, gluons, which are the carrier
of the color force. When QCD is dimensionally reduces to $1+1$ dimensions
[8,9,10], the gauge fields may be eliminated using their equations of
motion. What remains is the effectively long-range quark-quark force, which
is given by a linear two body potential [11]. This linear potential can be
understood in other ways. Since gluons are massless, their propagator in
momentum space is $1/q^{2}$. This is the same as the photon propagator,
which corresponds to the $\frac{1}{\left| r-r^{\prime }\right| }$Coulomb
potential. An appropriate wave function for the valance quarks inside the
proton should satisfy certain conditions. According to the boundary
condition of the problem, the quarks are considered as free particles in the
central region of the nucleon and they are tightly bounded at large
distance. Consequently, the nucleon space is classified into two regions.
The central part, where $s<s_{i}<s_{0}$ , the quark-wave function is taken
as plane wave. And in the peripheral region $s_{i}>s_{0}$, the quark wave
function is approximated to harmonic oscillator wave function. Here $s_{0}$%
is a parameter which specifies the boundary of the central region. In the
approximation of the straight line trajectories [3], the total phase shift
function may be expressed as the sum of the phase shift functions due to the
3- valence quarks and N- dipoles, as scattering centers,

\begin{equation}
\Xi (\stackrel{\_}{b},\{\stackrel{\_}{b}_{i}\})=\sum_{i}\xi _{i}(\stackrel{\_%
}{b}-\stackrel{\_}{b}_{i})+\sum_{i}^{dip}\xi _{i}(\stackrel{\_}{b}-\stackrel{%
\_}{b}_{i})
\end{equation}

$\qquad \xi _{i}$ is the phase shift due to the ith valence quark, and $\xi
_{i}^{dip}$is the phase shift function due to the i$^{th}$ electric dipole.
Defining a profile function as $\Gamma (\stackrel{\_}{b},\{\stackrel{\_}{b}%
_{i}\})$,

\begin{equation}
\Gamma (\stackrel{\_}{b},\{\stackrel{\_}{b}_{i}\})=1-\exp [i\Xi (\stackrel{\_%
}{b},\{\stackrel{\_}{b}_{i}\})]
\end{equation}

$\qquad \qquad $

Which could be expressed in terms of the two-body profile function $\Gamma
_{i}$,

\begin{eqnarray}
\Gamma (\stackrel{\_}{b},\{\stackrel{\_}{b}_{i}\}) &=&1-\prod [1-\Gamma _{i}(%
\stackrel{\_}{b}-\stackrel{\_}{b}_{i})]\prod [1-\Gamma _{i}^{dip}(\stackrel{%
\_}{b}-\stackrel{\_}{b}_{i})]  \nonumber \\
&=&(\sum_{j}^{3}\Gamma _{i}+\sum_{j}^{N}\Gamma _{j}^{dip})-(\sum_{j\neq
i}^{3}\Gamma _{i}\Gamma _{j}+\sum_{j\neq i}\Gamma _{j}^{dip}\Gamma
_{i}^{dip}+\cdot \cdot \cdot  \\
&&+(-1)^{N-1}\prod_{j}^{N}\Gamma _{j}^{dip}\prod_{i}^{3}\Gamma _{i} 
\nonumber
\end{eqnarray}

The terms of the first bracket in the R.H.S. of Eq.(2.4) represent the single
scattering or the impulse approximation, the second bracket is the
correction due to the binary collisions and finally the last term is the
correction due to the collision with all the constituent of the proton bag,
i.e. the collision at which all the three valence quarks (uud) and the N-
pairs contribute the reaction. Assuming that all the quark dipoles are
similar and equally probable, then Eq. (2.4) becomes,

\begin{eqnarray}
\Gamma (\stackrel{\_}{b},\{\stackrel{\_}{b}_{i}\}) &=&(\sum_{j}^{3}\Gamma
_{i}+N\Gamma ^{dip})-(\sum_{j\neq i}^{3}\Gamma _{i}\Gamma _{j}+\frac{N(N-1)}{%
2}(\Gamma ^{dip})^{2} \\
&&+\cdot \cdot \cdot +(-1)^{N-1}(\Gamma ^{dip})^{N}\prod_{i}^{3}\Gamma _{i} 
\nonumber
\end{eqnarray}

At sufficiently high energy, it is expected that the number N of created is
much greater than the number of valence quarks. And if this number is
linearly proportional to the center of mass energy Ecm of the reaction, then
the terms of Eq. (2.6) are energy dependent, and read as,

\begin{equation}
\Gamma (b,\{b_{i}\})\simeq E_{cm}\Gamma ^{dip}-\frac{E_{cm}(E_{cm}-1)}{2}%
(\Gamma ^{dip})^{2}+\cdot \cdot \cdot
\end{equation}

So that the first term in Eq. (2.6) is linearly dependent on $E_{cm}$ while
the second term that represents the double scattering is quadratically
dependent. This term is not important even at high energy, since the
electromagnetic fine structure constant $\alpha $ appears of second order
through the profile function that depresses the term as a whole. The result
of Eq.(2.6) is very important, since it explains clearly the linear
proportionality of the e-p total cross-section with the center of mass
energy.

The problem now is reduced to the two-body collision, in other words, we
have to formulate, only, for the simple two-body function and in terms of
the scattering potential,

\begin{equation}
\Gamma _{j}(\stackrel{\_}{b}-\stackrel{\_}{b}_{j})=1-\exp [-\frac{i}{2k(1-x)}%
\int_{-\infty }^{\infty }U(\stackrel{\_}{b}-\stackrel{\_}{b}_{j},\stackrel{\_%
}{z})dz]
\end{equation}

Assuming a Coulomb like potential field with the form,

\begin{equation}
U=\frac{U_{0}}{r}\exp (-\alpha r){ \qquad and \quad }\stackrel{\_}{r}=(%
\stackrel{\_}{b}-\stackrel{\_}{b}_{j})+\stackrel{\_}{z}
\end{equation}

\begin{equation}
\Gamma _{j}(\stackrel{\_}{b}-\stackrel{\_}{b}_{j})=1-\exp [\frac{iU_{0}}{%
2k(1-x)}K_{o}(\alpha (b^{2}+b_{j}^{2}-2bb_{j}\cos \phi _{j}))]
\end{equation}

$K_{o}(b^{2}+b_{j}^{2}-2bb_{j}\cos \phi _{j})$ is the modified Bessel
function or the so called Mc-Donald function of order zero. $U_{o}$, being
the strength of the field. For the two flavors u and d of the valence quarks
having fractional charges $2/3$ and $-1/2$ respectively, we get

\begin{equation}
\Gamma _{u}(\stackrel{\_}{b}-\stackrel{\_}{b}_{j})=1-\exp [-\frac{iU_{0}}{%
2k(1-x)}K_{o}(\frac{2}{3}\alpha (b^{2}+b_{j}^{2}-2bb_{j}\cos \phi _{j}))]
\end{equation}

And

\begin{equation}
\Gamma _{d}(\stackrel{\_}{b}-\stackrel{\_}{b}_{j})=1-\exp [-\frac{iU_{0}}{%
2k(1-x)}K_{o}(-\frac{1}{2}\alpha (b^{2}+b_{j}^{2}-2bb_{j}\cos \phi _{j}))]
\end{equation}

To calculate the profile function $\Gamma _{j}^{dip}(\stackrel{\_}{b}-%
\stackrel{\_}{b}_{j})$due to the scattering from a single dipole, let us
consider that a distance $\stackrel{\_}{\ell }$ separates the two quarks of
the dipole. The potential field due to the dipole is given by,

\begin{equation}
U_{dip}=-\frac{\lambda }{\ \left| \stackrel{\_}{r}-\stackrel{\_}{r}_{dip}+%
\stackrel{\_}{\ell }/2\right| }+\frac{\lambda }{\ \left| \stackrel{\_}{r}-%
\stackrel{\_}{r}_{dip}-\stackrel{\_}{\ell }/2\right| }
\end{equation}

And putting, $\stackrel{\_}{r}=\stackrel{\_}{b}+\stackrel{\_}{z}$, $%
\stackrel{\_}{r_{dip}}=\stackrel{\_}{b_{dip}}+\stackrel{\_}{z}$ , $\stackrel{%
\_}{\ell }=\stackrel{\_}{b_{\ell }}+\stackrel{\_}{z}$ where $\stackrel{\_}{z}
$ is the projection along the collision axis and b is the projection in the
impact parameter plane. So that the phase shift function due to a dipole is,

\begin{equation}
\Xi ^{dip}(b,b_{dip},\ell )=\frac{\lambda }{(k(1-x)}\log [\frac{(\stackrel{\_%
}{b}-\stackrel{\_}{d}_{dip}+0.5\stackrel{\_}{b_{\ell }})^{2}}{(\stackrel{\_}{%
b}-\stackrel{\_}{d}_{dip}-0.5\stackrel{\_}{b_{\ell }})^{2}}]
\end{equation}

and the total phase shift due to the N-dipoles is,

\begin{equation}
\Xi _{tot}^{dip}(b,b_{dip},\ell )=\frac{\lambda }{(k(1-x)}\sum_{j}^{N}\log [%
\frac{(\stackrel{\_}{b}-\stackrel{\_}{b}_{j}+0.5\stackrel{\_}{b_{\ell }})^{2}%
}{(\stackrel{\_}{b}-\stackrel{\_}{b}_{j}-0.5\stackrel{\_}{b_{\ell }})^{2}}]
\end{equation}

Then Eq.(2.1) becomes,

\begin{equation}
f_{ms}(q,x)=\frac{ik(1-x)}{2\pi }\int d^{2}\stackrel{\_}{b}\exp (i\stackrel{%
\_}{q}\cdot \stackrel{\_}{b})\left\langle \Psi \{\stackrel{\_}{s}%
_{i}\}\left| \Gamma (\stackrel{\_}{b},\{\stackrel{\_}{s}_{i}\}\right| \Psi \{%
\stackrel{\_}{s}_{i}\}\right\rangle
\end{equation}

\section{ Results and Discussion}

The model predictions are tested with the experimental data of SLAC-E-049,
SLAC-E-061, and SLAC-E-089 [12] that concerns the deep inelastic scattering
in the lab momentum ranges \{3.748 - 24.503\}. A spherical proton bag is
assumed with root mean square radius $R_{rms}=1.2$ $fm$. Since the quarks
are almost considered independent, then the proton spatial wave function is
taken as, $\Psi _{p}=\prod_{u,u,d}\Phi _{i}$ , with $\Phi _{i}$ as a plane
wave for $s<s_{o}$ and $\Phi _{i}$ is a harmonic oscillator wave function
for $s>s_{o}$. The parameter $s_{o}$ is adjusted to fit with the $e-p$
inelastic cross-section at the corresponding center of mass energy. It is
found that $s_{o}$ is $\frac{1}{3}R_{rms}$ at the considered energy. The
total profile function of the e-p scattering is calculated with a weight
factor corresponding to the quark probability density inside the proton bag.

Fig.(1) shows the individual terms of the average profile function
corresponding to the single, double and the triple scattering. Of course the
first term is the most important one allover the impact parameter range. The
double scattering term represents about 0.02 from the first one at the
central region. This ratio is greater than the reciprocal of the fine
structure constant $\frac{1}{137}$. The central region is populated with
high density of states, so there will be a good chance for the double
collision to occur. The values of the triple scattering may be neglected. In
other words, the shadowing has significant effect only up to the double
scattering. The resultant profile function is plotted in Fig.(2), which
shows slow decrease with the impact parameter $b$, reflecting the Coulomb
behavior of the assumed scattering potential. The ratio of the imaginary to
the real part of the profile function is an important coefficient. It is
just a measure of the degree of inelasticity of the reaction. As
demonstrated in Fig.(3), the imaginary to the real part of the coefficient
shows that the central region is characterized by the inelastic scattering.
The inelasticity decreases with the impact parameter i.e. toward the
peripheral region. The total differential cross-section is calculated
according to Eq.(2.15) at $x=0.5$, and incident energy $14GeV$. The prediction
of the multiple scattering model, Fig.(4) shows a large peak at small values
of $q^{2}$, followed by several humps of successively decreasing height. The
dashed line represents the $1/q^{4}$ behavior of the point particle
scattering. It seems to tangents the humps of the multiple scattering model.
It is expected that the two curves coincide at low energies where the
electron wavelength is long enough and cannot probe the microscopic
structure of the proton. Now replacing the factor $1/q^{4}$ that appears in
Eq.(1.1) by the corresponding multiple scattering form $\left| f_{ms}\right|
^{2}$, as in Eq.(2.15), and then recalculate the enhanced structure function $%
F_{2}^{Enh}$. Fig.(5-7) show the model prediction at $x=0.08,0.125$ and $0.55
$ respectively, compared with the prediction of the parton model $F_{2}$
[13,14]. Appreciable deviation is observed at small values of $x$, which
correspond to most deep inelastic scattering. As $x$ increases, the two
models come closer, where the reaction is directed toward the peripheral
region that is characterized by single collisions only. In spite of the
observed variation in the structure function with $q^{2}$, but it still
keeps its own characteristics. In other words, it leads to the half spin
value of quarks forming the protons and also it doesn't conflict with the
values of their fractional charge.

\begin{figure}[thb]
\centerline{
   \epsfig{figure=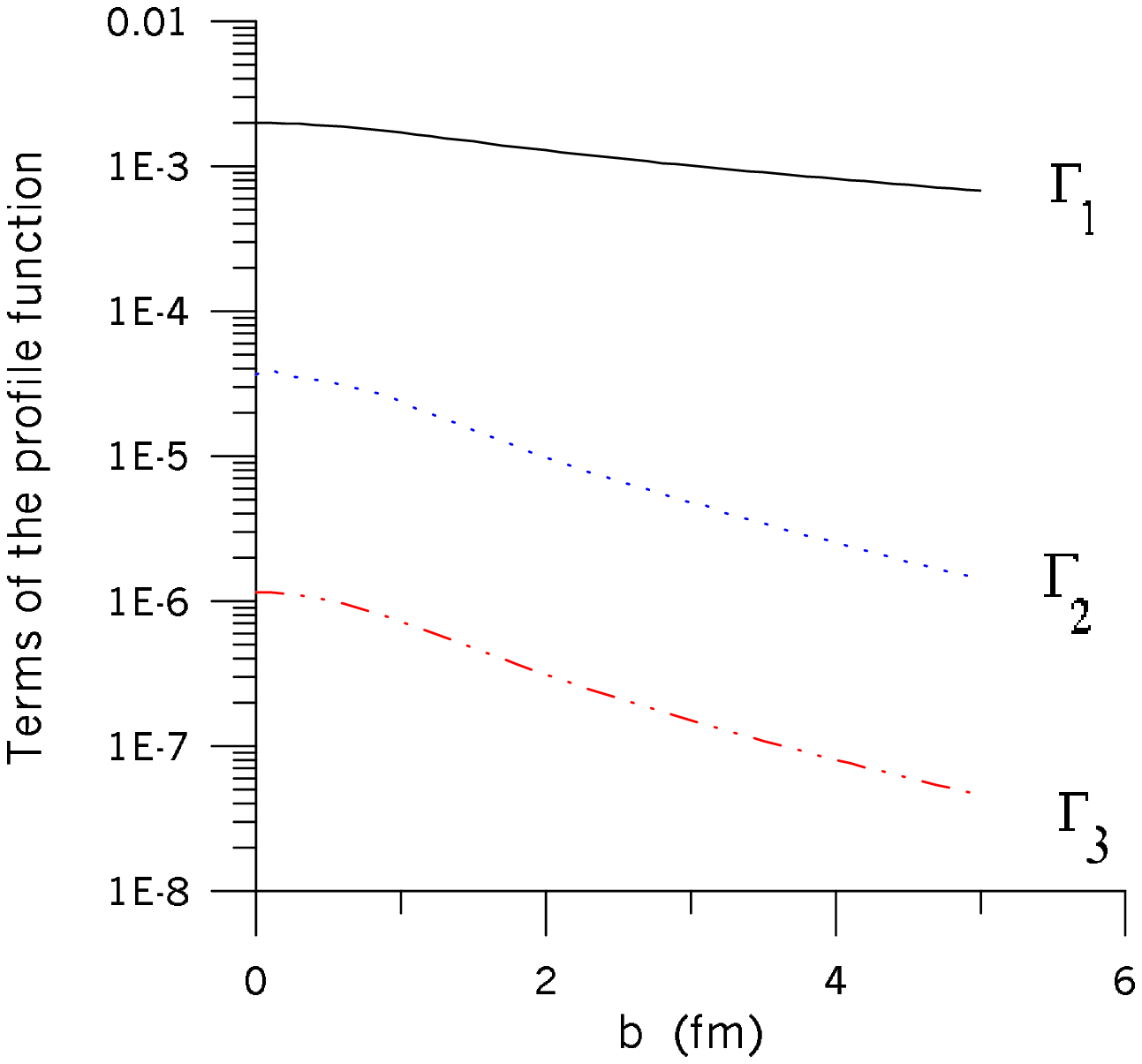,width=0.4\textwidth,clip=}
           }
\caption{
The first three terms of the profile
function for the e-p multiple scattering. The impulse approximation, binary 
collision and the third order term.}
\end{figure}

\begin{figure}[thb]
\centerline{
   \epsfig{figure=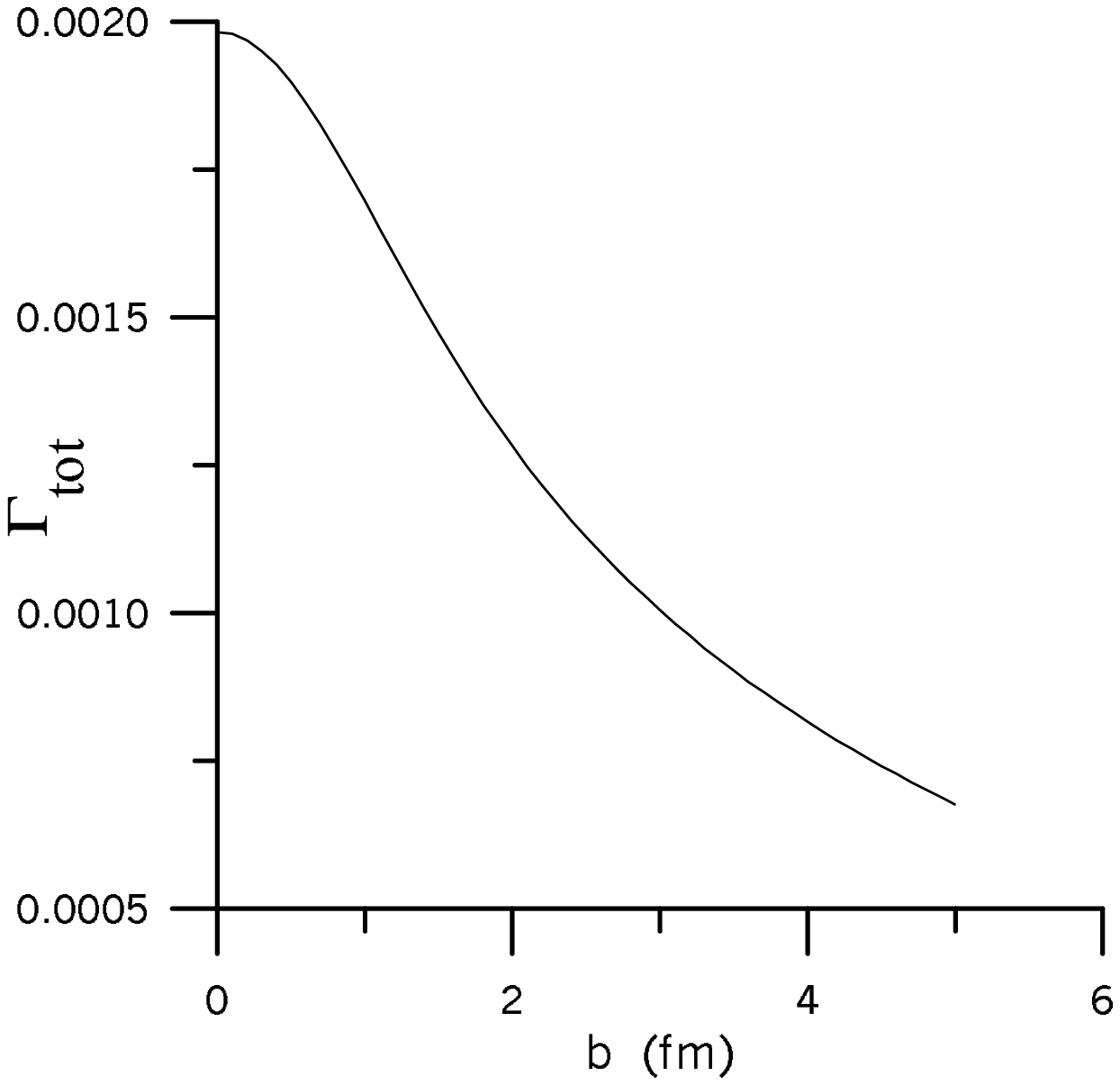,width=0.4\textwidth,clip=}
           }
\caption{
The total profile function of the Coulomb
field due to the interaction of the incident electron with the 3-valance quarks 
forming the target proton.}
\end{figure}

\begin{figure}[thb]
\centerline{
   \epsfig{figure=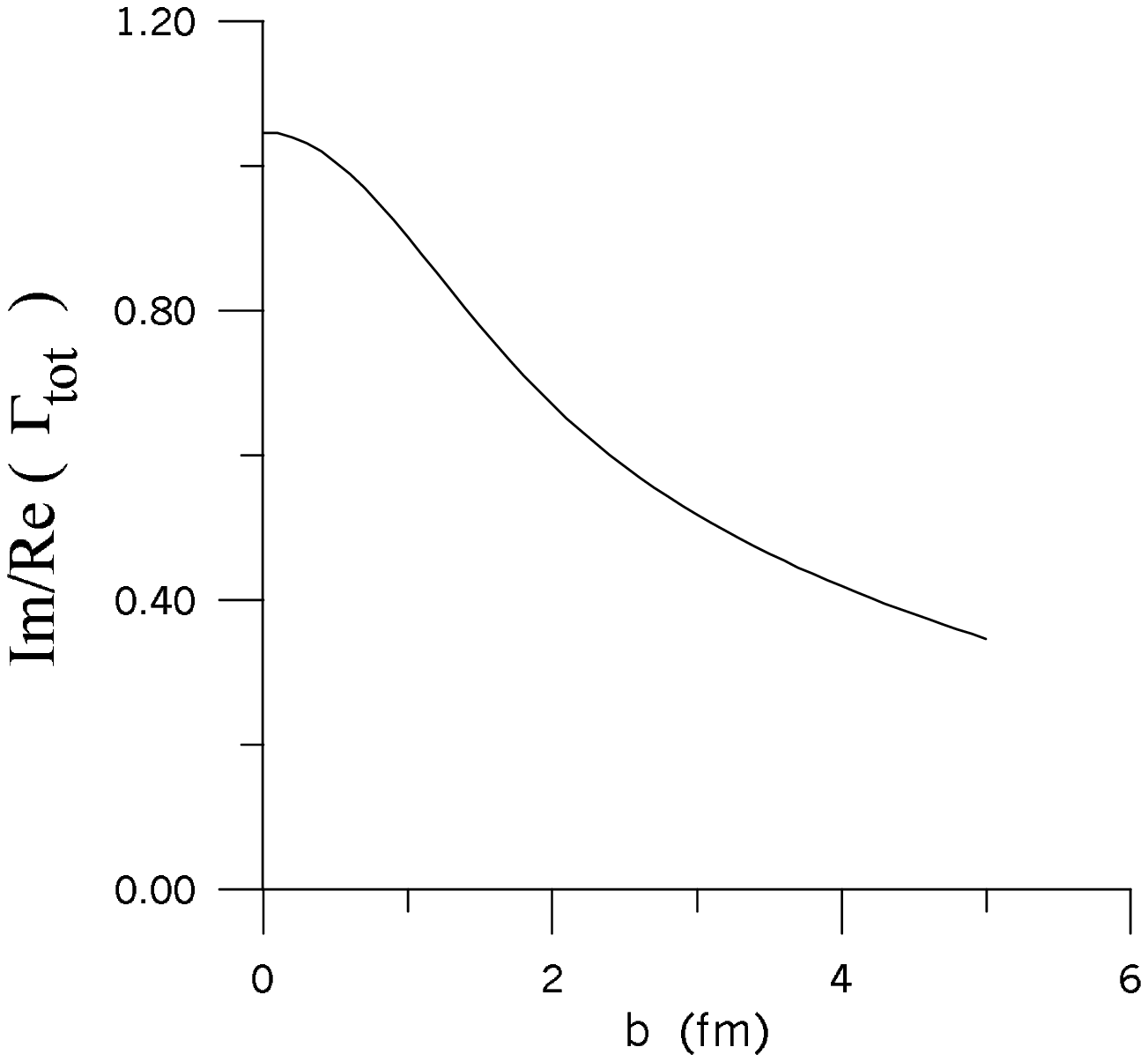,width=0.4\textwidth,clip=}
           }
\caption{
The ratio of imaginary to real part of the
total profile function of the Coulomb field due to the interaction of the 
incident electron with the 3-valance quarks forming the target proton.}
\end{figure}

\begin{figure}[thb]
\centerline{
   \epsfig{figure=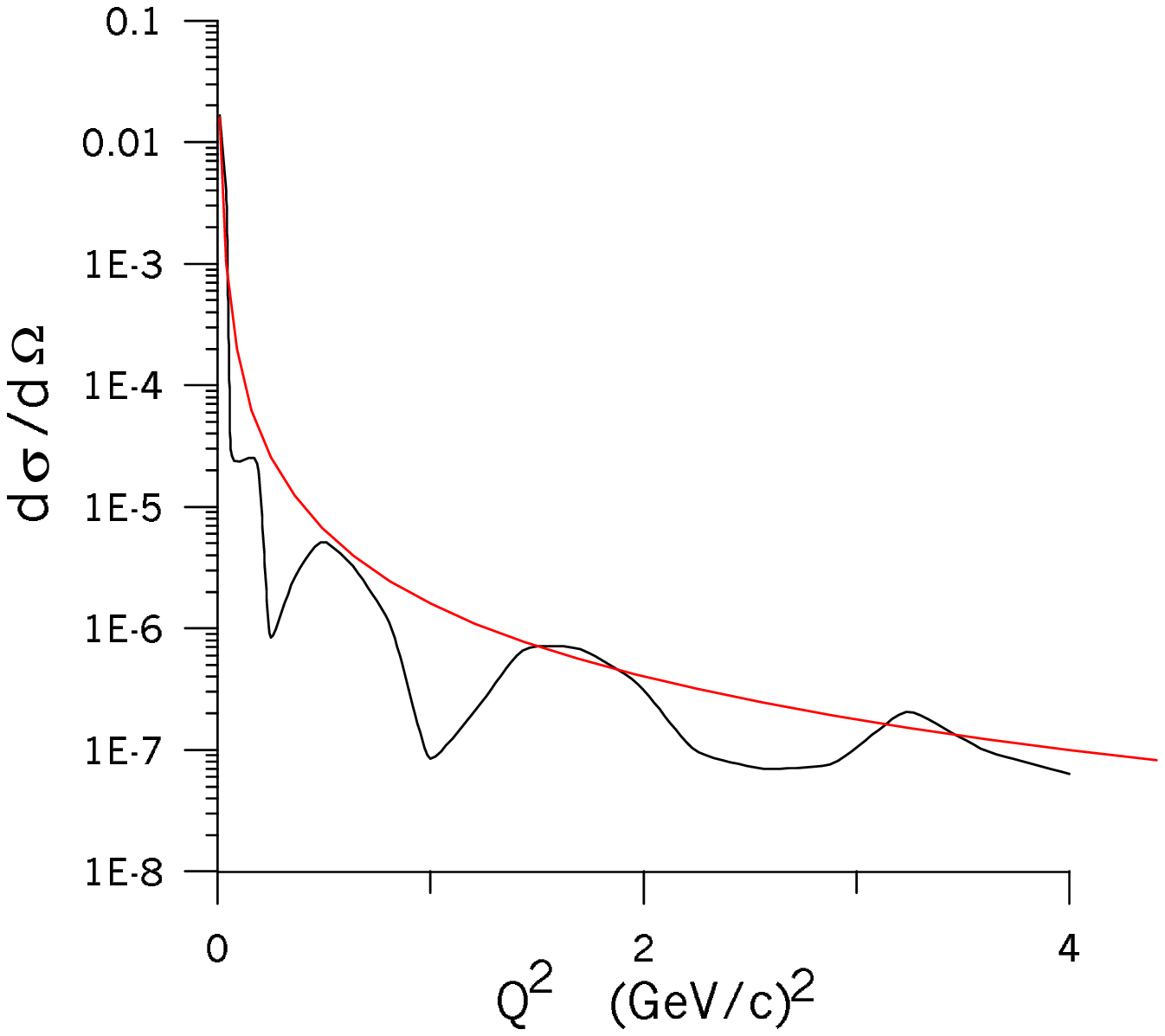,width=0.4\textwidth,clip=}
           }
\caption{
The differential cross section  for the
scattered electron in e-p experiment at 14 GeV. The black line represents the 
prediction of the multiple scattering model and the dotted red one is the 
$1/q^{4}$ law.}
\end{figure}

\begin{figure}[thb]
\centerline{
   \epsfig{figure=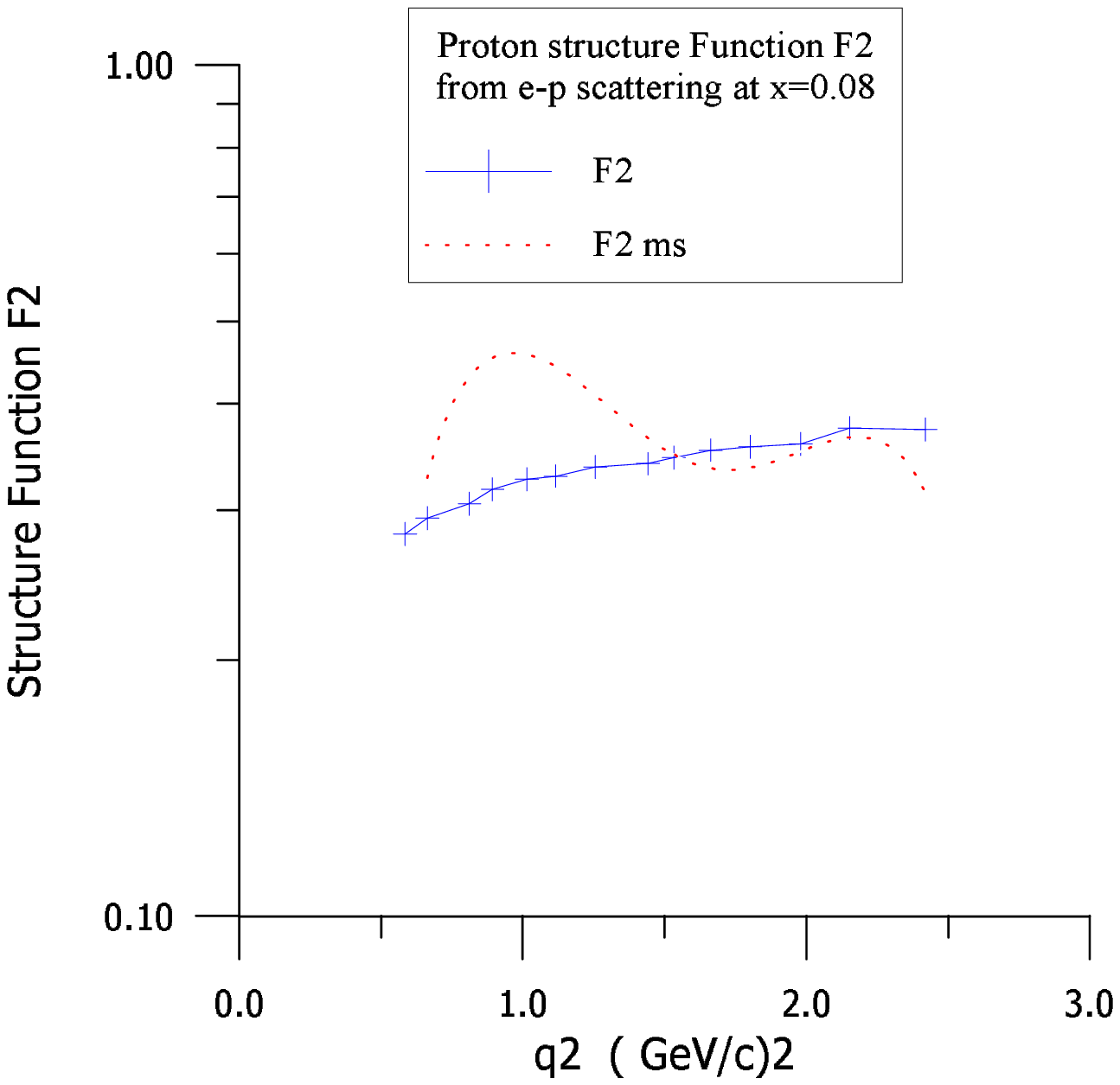,width=0.4\textwidth,clip=}
           }
\caption{
The proton structure function as deduced
from the experiments SLAC-E-049 and SLAC-E-089 at x = 0.08. The dashed line 
represents the modified function due to multiple scattering.}
\end{figure}

\begin{figure}[thb]
\centerline{
   \epsfig{figure=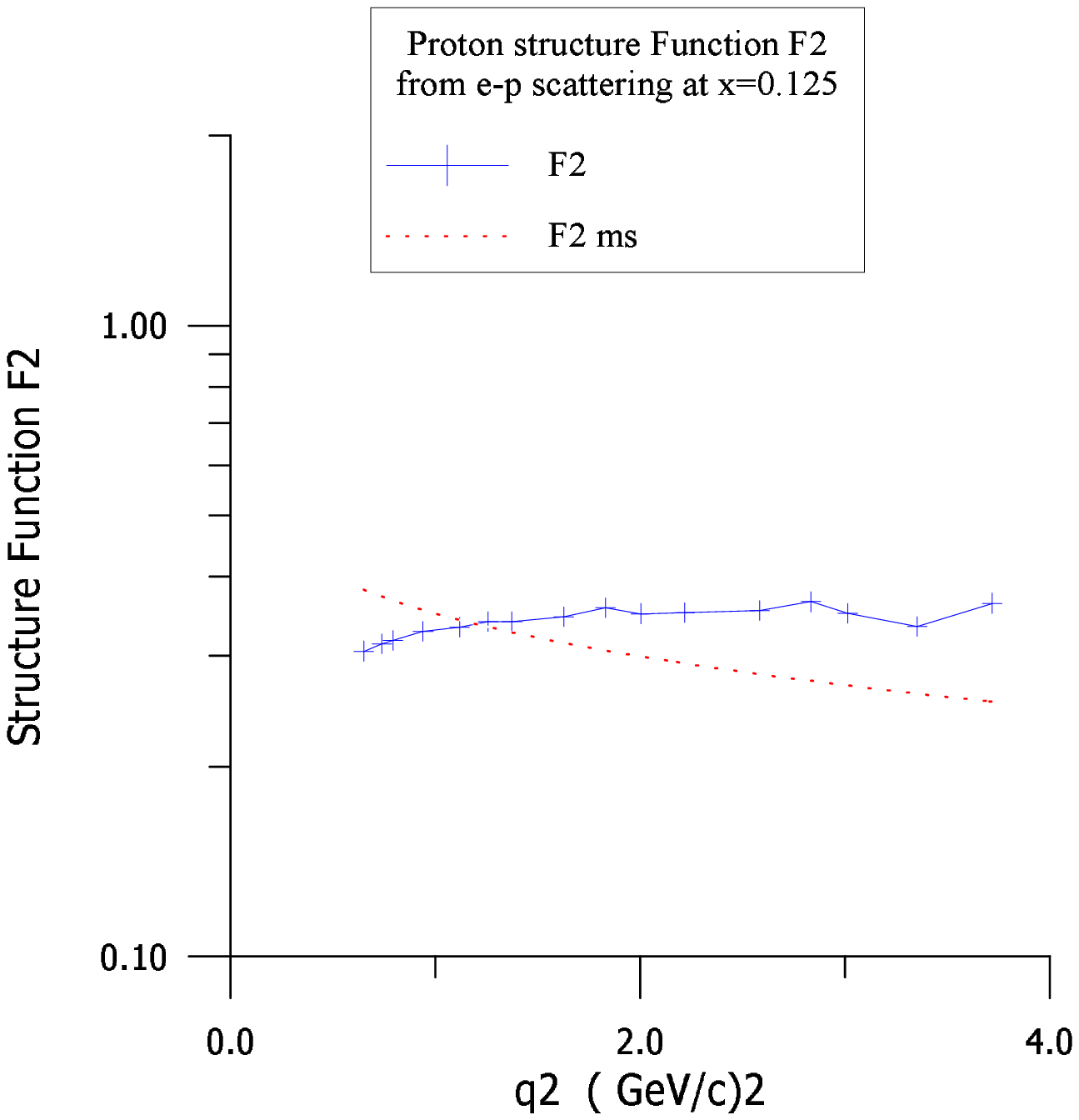,width=0.4\textwidth,clip=}
           }
\caption{
The proton structure function as deduced
from the experiments SLAC-E-049 and SLAC-E-089 at x = 0.125. The dashed line 
represents the modified function due to multiple scattering.} 
\end{figure}

\begin{figure}[thb]
\centerline{
   \epsfig{figure=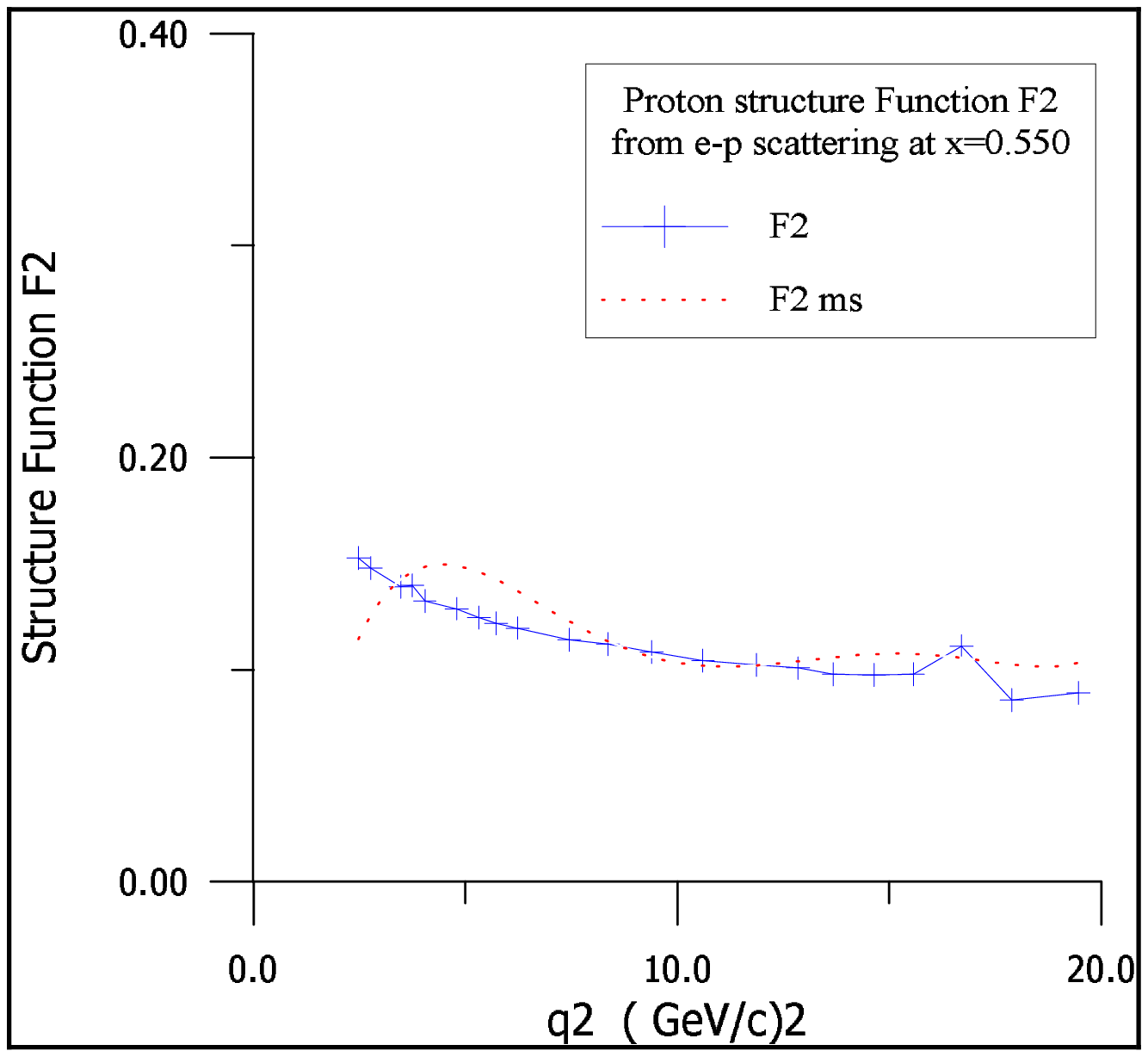,width=0.4\textwidth,clip=}
           }
\caption{
The proton structure function as deduced
from the experiments SLAC-E-049 and SLAC-E-089 at x = 0.550. The dashed line 
represents the modified function due to multiple scattering.}
\end{figure}

\section{Concluding Remarks.}

\begin{itemize}
\item  The multiple scattering of the electron with constituent of the
proton is very important to estimate its structure function.

\item  The number of scattering centers inside the proton bag plays an
important role to explain the linear increase of the total cross-section of
the electron-proton scattering with the center of mass energy.

\item  Terms up to the double scattering shows significant shadowing effect
during the collision.

\item  The square of the scattering amplitude shows a peak followed by humps
of decreasing height. The asymptotic behavior of which has the form of 1/q4.

\item  Appreciable deviation to the structure function is due to the
multiple scattering effect, particularly for the low values of the Bjorken
scaling variable x which correspond to the deep inelastic scattering.
\end{itemize}

\end{document}